# Terahertz channel power and BER performance in rain

YUHENG SONG,[1] JIAYUAN CUI,[1] GUOHAO LIU,[1] JIABIAO ZHAO,[1] MINGXIA ZHANG,[1] JIACHENG LIU,[1] DA LI,[1] PEIAN LI,[1] CHEN YAO,[1,*] FEI SONG,[2] HONG LIANG,[3] JIANJUN MA[1,*]

[1]*Beijing Key Laboratory of Millimeter and Terahertz Wave Technology, Beijing Institute of Technology, Beijing, 100081, China*
[2]*School of Electronic and Information Engineering, Beijing Jiaotong University, Beijing, 100044 China*
[3]*Radar Operation Control Department, China Meteorological Administration, Beijing, 100044, China*
**Corresponding: Jianjun Ma: jianjun_ma@bit.edu.cn*
**Corresponding: Chen Yao: chenyao@bit.edu.cn*

**Abstract:** Terahertz (THz) communications have emerged as a promising technology for 6G networks due to their potential for achieving terabit-per-second data rates. However, the impact of rainfall on THz channel characteristics remains incompletely understood, particularly regarding power attenuation mechanisms and bit error rate (BER) performance. This article presents a systematic measurement-based and theoretical investigation of line-of-sight (LoS) THz channel behavior under rainfall conditions, methodically examining both power attenuation mechanisms and bit error rate (BER) performance. Our experimental campaign, conducted at frequencies of 220-230 GHz over a 54-meter outdoor channel, is complemented by analytical frameworks incorporating ITU-R and Mie scattering models. The study reveals that while rain induces significant power attenuation, multipath scattering effects remain minimal, with Rician K-factors maintaining high values. Notably, we observe substantial variations in power loss under constant rain rates, attributed to dynamic changes in raindrop size distribution. Comparative analysis demonstrates superior BER performance of Quadrature Amplitude Modulation (QAM) in rainfall conditions, while revealing increased environmental sensitivity at higher frequencies. These findings underscore the necessity for adaptive modulation schemes and strategic frequency planning in future THz communication systems.

## 1. Introduction

The rapid evolution of wireless communication technologies, driven by emerging applications such as extended reality (XR), artificial intelligence (AI), and massive Internet of Things (IoT), has led to unprecedented demands on wireless network capacity and performance [1-4]. While current 5G systems operate primarily in sub-6 GHz and millimeter-wave bands, achieving the ambitious performance targets of 6G networks—including 1 Tbps peak data rates, 10 Gbps user-experienced rates, and sub-millisecond latency [5]—necessitates exploration of higher frequency bands, particularly the terahertz (THz) spectrum (0.1-10 THz) [6]. THz communications offer several inherent advantages, as ultra-wide bandwidths supporting Tbps data rates [7], enhanced security through high directivity [8], and reduced interference due to high atmospheric absorption [9, 10]. These characteristics position THz spectrum as a promising candidate for various 6G applications, including ultra-high-speed wireless backhaul, sensing-integrated communications, and ultra-dense networks [5, 11, 12]. However, the practical deployment of THz systems faces significant challenges, particularly in outdoor environments where atmospheric effects can severely impact channel performance [13-16]. Among these, rainfall presents one of the most significant challenges for THz channel propagation due to complex interaction mechanisms, statistical variability and temperature and humidity coupling [17].

The study of THz channel propagation through rainfall has evolved significantly over the past decade. Early work focused primarily on empirical measurements and simple analytical models. Ishii et al. conducted pioneering measurements at frequencies up to 355 GHz, demonstrating attenuation levels of approximately 6 dB/km at a rain rate of 10 mm/hr [18-20]. These findings were later expanded by Hirata et al., who investigated rain effects on a 120 GHz wireless link, revealing that conventional prediction models often underestimate the BER degradation in heavy rainfall [21], due to the lack of channel fading. More recent studies have employed sophisticated measurement techniques, including THz time-domain spectroscopy, to characterize both amplitude and phase effects of rain on THz channels [13, 22]. On the other hand, the modeling of rain effects on THz propagation has traditionally followed two main approaches: empirical models based on the ITU-R recommendations and theoretical frameworks using Mie scattering theory. The ITU-R model, while widely adopted for its simplicity, has shown limitations in accurately predicting attenuation at THz frequencies, particularly in tropical regions with intense rainfall [23, 24]. Several publications have proposed modifications to improve its accuracy. Norouzian and Li developed a corrected model incorporating measured drop size distributions, achieving better agreement with experimental data at frequencies up to 450 GHz [23, 25]. Similarly, Islam et al. introduced a hybrid approach combining the ITU-R model with local meteorological data, demonstrating improved prediction accuracy in various climatic zones [26]. Mie scattering-based approaches offer more rigorous theoretical treatment but face challenges in practical implementation. These models require accurate characterization of raindrop size distributions, which can vary significantly with geographical location and weather conditions [27]. Various distribution models have been proposed, including the Marshall-Palmer (M-P), Laws-Parsons (L-P), and Weibull distributions [19, 28]. A recent publication has shown that no single distribution model adequately describes all rainfall conditions, suggesting the need for adaptive modeling approaches [25].

Despite these advances, several crucial aspects of THz channel propagation in rain remain inadequately addressed. Most existing studies (as shown in Table 1) focus on power attenuation mainly, while the influence of rain-induced fading on power profile and bit error performances remains poorly understood, particularly for advanced modulation schemes. Ma et al. demonstrated that rainfall introduces fast fading from individual raindrop interactions, besides slow fading due to average power attenuation [29], but they did not characterize that further.

**Table 1. Typical investigations on THz channel performance in rain.**

| Freq. (GHz) | Distance | Investigation method | Modulation scheme | Ref. |
|---|---|---|---|---|
| 100-500 | 1 km | Theoretical investigation | Un-modulated | [30] |
| 100–1000 | 4 m | Indoor measurement | Un-modulated | [31] |
| 313 | 1 km | Outdoor measurement | Un-modulated | [19] |
| 355.2 | 33 m | Outdoor measurement | Un-modulated | [18] |
| 173, 230 | 1 km | Outdoor measurement | Un-modulated, power profiles | [32] |
| 100, 1000 | ≤865 m | Outdoor measurement | Un-modulated | [33] |
| 103 | 390 m | Outdoor measurement | Un-modulated | [34] |
| 300 | 160 m | Outdoor measurement | Un-modulated | [23] |
| 100-1000 | < 1 km | Theoretical investigation | Un-modulated, power profiles | [35] |
| 130-150 | 70 m | Outdoor measurement | Un-modulated | [9] |
| 100-1000 | 100 m | Theoretical investigation | Un-modulated | [36] |
| 1000 | 30-200 m | Theoretical investigation | Un-modulated | [37] |
| 140 | 41.4 m | Outdoor measurement | Un-modulated | [25] |
| 120 | 400 m | Outdoor measurement | ASK modulation | [21] |
| 275, 400 | ≤12 km | Theoretical investigation | X-PSK and X-QAM modulations | [38] |
| 162 | 1 m | Indoor measurement | 16-QAM modulation | [13] |
| 625 | 3 m | Indoor measurement | OOK modulation | [31] |

This article tries to present a comprehensive experimental and theoretical investigation of THz power and BER performance in rainfall conditions, conducted in the passageway between Wencui Buildings I and F at Beijing Institute of Technology's Liangxiang campus. The remainder is organized as follows: Section II details the experimental setup and rainfall environment characterization; Section III analyzes the channel power distribution, multipath effects, and temporal stability under rainfall, together with rain attenuation prediction methodology; Section IV evaluates modulation scheme performance in fading channels under rainfall and provides comprehensive BER analysis under varying environmental conditions; and Section V concludes the article with key findings and future research directions.

## 2. Channel measurement setup

The experimental apparatus comprised a comprehensive terahertz transmission and reception system, as shown in Fig. 1. The transmitter chain consisted of a Ceyear 1465D signal generator producing baseband signals (100 kHz - 20 GHz) with the power of 0 dBm, followed by a Ceyear 82406D frequency multiplier module with an ×18 multiplication factor for up-conversion to the desired terahertz frequencies (220-325 GHz). A TCNPA-220 power amplifier was integrated between the frequency multiplier and the antenna system to enhance signal strength. The antenna configuration incorporated a Ceyear 89901S horn antenna coupled with a high-density polyethylene (HDPE) lens featuring a 300 mm focal length. This integrated transmission system achieved a remarkable composite gain of approximately 60.5 dB, with individual contributions of 17.5 dB from the lens, 23 dB from the antenna, and 20 dB from the power amplifier. The receiver architecture mirrored the antenna configuration of the transmitter, utilizing an identical Ceyear 89901S horn antenna and HDPE lens combination, coupled to a Ceyear 71718 power sensor for direct signal detection. Both transmitter and receiver assemblies were precisely positioned at a height of 162.5 cm above ground, strategically exceeding the first Fresnel zone radius (calculated as 58.3 cm for 220 GHz). The system maintained a consistent beamwidth of 6.7° across all experimental frequencies. To ensure equipment protection while maintaining measurement integrity, both transmitter and receiver units were sheltered under building eaves, minimizing direct exposure to precipitation while allowing unobstructed channel propagation through the rain-affected region. The measurement protocol was implemented through a computerized control and data acquisition system. A laptop computer orchestrated the signal generator operations and power sensor measurements through automated software control, maintaining a consistent sampling rate of 7 Hz throughout the experimental campaign. The Ceyear-based measurement system represents the sole terahertz channel measurement platform available in our laboratory, with our facility being one of the few research institutions in northern China equipped with this specific configuration for terahertz experimentation.

The measurements were conducted through an outdoor experimental campaign on July 24, 2024, employing a fixed point-to-point THz channel configuration. The experimental setup was strategically positioned in the passageway between Wencui Buildings I and F at the Liangxiang Campus of Beijing Institute of Technology, selected for its optimal atmospheric research conditions. The channel configuration maintained a total path length of 54 meters, with 41.5 meters directly exposed to rainfall conditions. Measurements were performed at three distinct frequencies - 220 GHz, 225 GHz, and 229 GHz - chosen due to the bandwidth limitation of our TCNPA-220 power amplifier, which work in the frequency of 220 GHz-230 GHz only. The investigation encompassed both clear weather and rainfall conditions to establish comprehensive baseline and weather-impacted channel characteristics. The temporal measurement strategy was carefully designed to capture channel behavior across varying rainfall intensities. Measurements were conducted during a three-hour window from 19:00 to 22:00 local time. Individual channel measurements were precisely time-limited to one-minute duration, optimizing the acquisition of frequency-dependent characteristics under consistent rainfall conditions. This measurement duration aligns with established protocols for rainfall attenuation prediction [39] while enabling rapid sequential measurements across multiple frequencies under stable precipitation rates. This methodology facilitated the compilation of a robust dataset characterizing channel behavior across diverse weather conditions and frequencies. The experimental dataset was augmented through correlation with atmospheric measurements from the China Meteorological Administration (CMA) observation center. The proximity of the meteorological station to the Liangxiang Campus enabled precise characterization of local atmospheric conditions throughout the measurement campaign. This supplementary meteorological data provided crucial validation of experimental conditions and supported comprehensive analysis of THz channel performance under varying atmospheric parameters.

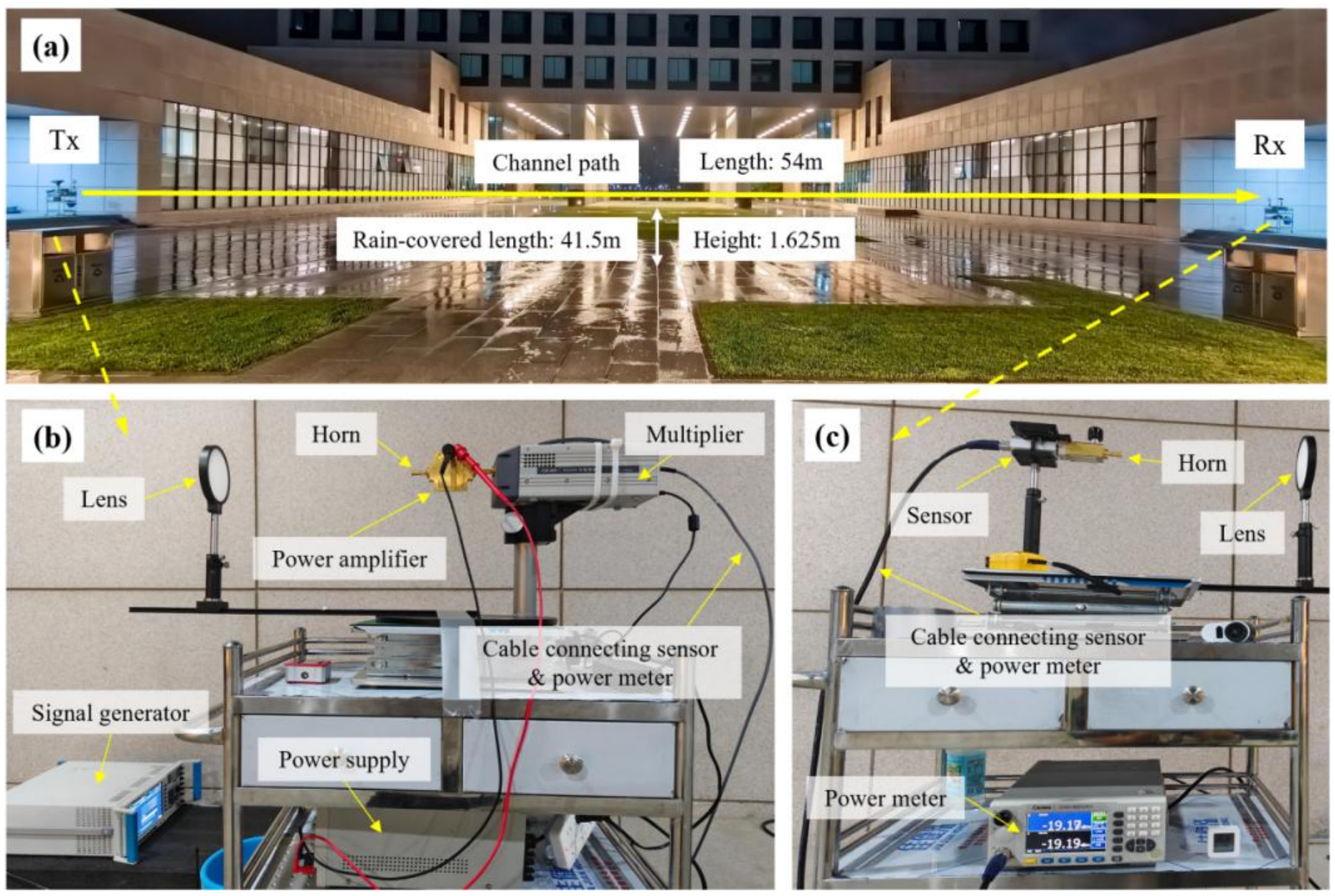

**Fig. 1.** THz channel measurement setup at the Liangxiang Campus of Beijing Institute of Technology (BIT). (a) The outdoor channel between Wencui Buildings I and F, with both the transmitter (Tx) and receiver (Rx) placed under the eaves; (b) Transmitter hardware; (c) Receiver hardware.

## 3. Channel power performance

In this section, we concentrate on performance of detected channel power, which is a pivotal metric for the design and assessment of wireless communication systems, under environmental conditions of approximately 27°C and 82% relative humidity (RH). In order to assess the performance of THz channels under two distinct atmospheric conditions, clear weather and rainy, we utilized the cumulative distribution function (CDF) to analyze the power measurement results. The data were collected over different time periods, thereby enabling a comparative analysis of the performances of terahertz channels under varying rain rates.

### *3.1. Experimental measurements*

The CDF curves of THz channels at different frequencies under different weather conditions are shown in Fig. 2. It can be seen that presence of rainfall makes the SNR decreases for all the channels. This is not surprising when compared with the observation that there is no obvious difference in the shapes of the CDF curve. This means there is no obvious multipath scattering phenomenon at the receiver side, which is out of our expectation and different with the channel measurements at 140 GHz in reference [9]. Thus, for further analysis, we take fitting to the measured data (see Fig. 2 and Table 2) by employing two distribution models - Rician and Weibull. Here, we did not consider the Stable distribution model, as there is no heavy-tailed phenomena observed [40]. We employed the coefficient of determination - R-Square - to evaluate the success of the fitted model where it ranges from - infinity to 1, where value one considers the optimum regression model to the measured data [41]. It was found that the Rician distribution model performed relatively better in fitting the data. This implies that while there is some multipath scattering, it is not as significant as might be expected, particularly given the lack of observable changes in the CDF shapes. This could mean that the THz channels might still be able to deliver reliable communication under rainy conditions, as the presence of a strong line-of-sight (LOS) component helps mitigate the impact of scattering and attenuation.

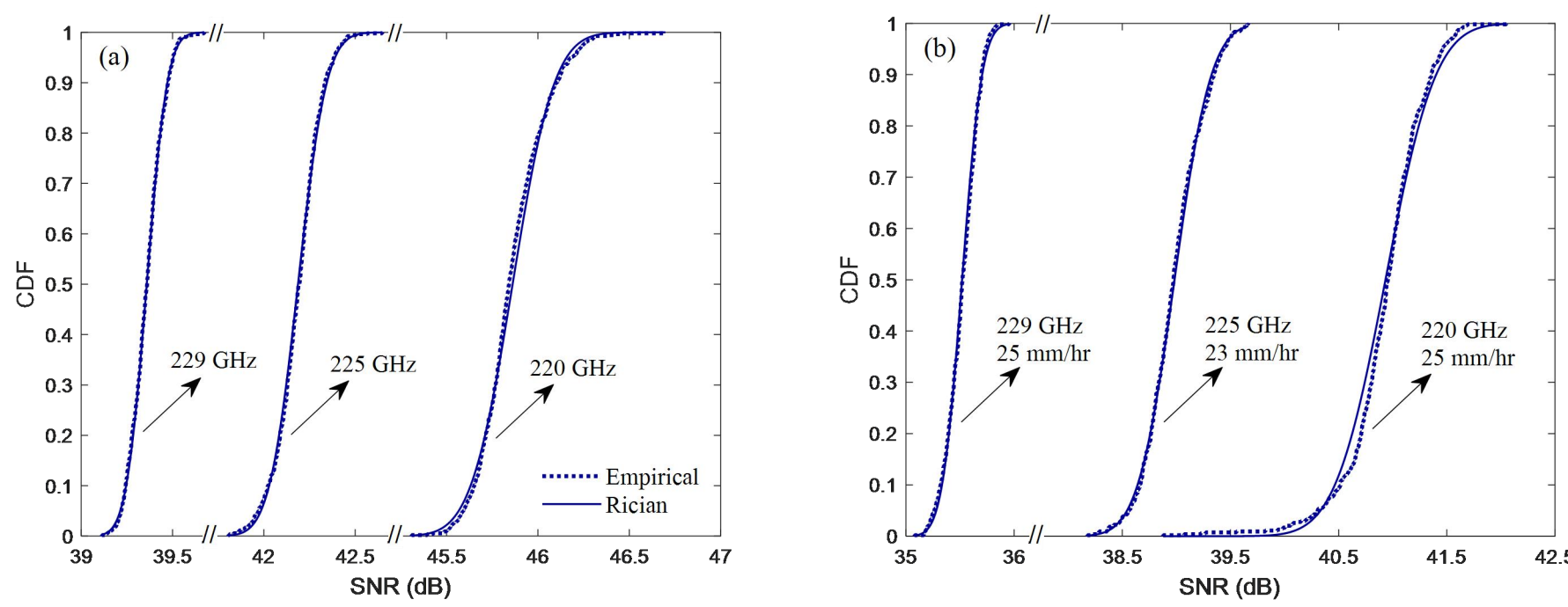

Fig. 2. CDF profile for received SNR (a) in clear weather and (b) in rain

**Table 2. Fitting parameters to the CDF profiles for the received SNR. (Parameter μ represents the mean value of the received power level, $\sqrt{\sigma^2}$ demotes the variance of the received signal's power level, K symbolizes the K-factor for the Rician distribution across different frequencies and weather conditions, $R^2$ can be used to reflect how well the model fits the measured data, while the scale and shape parameters define the specific characteristics of the Rician and Weibull distributions).**

| Frequency | Weather condition | $\mu$ | $\sqrt{\sigma^2}$ ($\times 10^{-8}$) | K (Ricain) | $R^2$ (Rician) | Scale parameter (Rician) | $R^2$ (Weibull) | Shape parameter (Weibull) | Scale parameter (Weibull) |
|---|---|---|---|---|---|---|---|---|---|
| 220 GHz | Clear | -16.85 | 2.5985 | 44.86 | 0.9967 | 0.1855 | 0.9785 | 46.02 | 216.84 |
| | Rain (8 mm/hr) | -16.96 | 1.1778 | 48.38 | 0.9953 | 0.1222 | 0.9961 | 45.42 | 401.35 |
| | Rain (10 mm/hr) | -17.71 | 0.5680 | 50.04 | 0.9991 | 0.0925 | 0.9931 | 41.66 | 452.45 |
| | Rain (25 mm/hr) | -17.96 | 2.6158 | 42.91 | 0.9981 | 0.2044 | 0.9929 | 40.53 | 210.37 |
| 225 GHz | Clear | -17.59 | 1.0392 | 47.65 | 0.9987 | 0.1235 | 0.9894 | 42.24 | 338.23 |
| | Rain (8 mm/hr) | -17.82 | 1.0278 | 47.26 | 0.9951 | 0.1259 | 0.9765 | 41.16 | 287.61 |
| | Rain (10 mm/hr) | -18.15 | 0.5416 | 49.38 | 0.9992 | 0.0950 | 0.9914 | 39.61 | 447.24 |
| | Rain (23 mm/hr) | -18.28 | 4.1560 | 40.26 | 0.9990 | 0.2673 | 0.9864 | 39.11 | 148.92 |
| | Rain (25 mm/hr) | -18.39 | 2.5260 | 42.20 | 0.9986 | 0.2111 | 0.9951 | 38.56 | 201.49 |
| 229 GHz | Clear | -18.20 | 0.5061 | 49.57 | 0.9997 | 0.0924 | 0.9908 | 39.39 | 420.51 |
| | Rain (6 mm/hr) | -18.45 | 0.8489 | 46.81 | 0.9998 | 0.1233 | 0.9909 | 38.27 | 312.36 |
| | Rain (10 mm/hr) | -19.18 | 0.2794 | 50.18 | 0.9983 | 0.0769 | 0.9799 | 35.17 | 435.00 |
| | Rain (23 mm/hr) | -19.23 | 4.0257 | 41.97 | 0.9960 | 0.2195 | 0.9976 | 35.06 | 129.20 |
| | Rain (25 mm/hr) | -19.09 | 1.0408 | 44.65 | 0.9985 | 0.1469 | 0.9954 | 35.57 | 256.74 |

It should also be noted from Table 2 that the variation of K-factor in clear and rainy weathers is not very much, even though there is an observable decreasing. The K-factor is a parameter that quantifies the ratio of the power in the direct LOS path to the power in the scattered paths (multipath components) [14, 42]. A higher K-factor indicates a stronger LOS channel path relative to the scattered paths, while a lower K-factor suggests more multipath interference. Although there is a decrease in the K-factor in rain, its stability implies that rain does not introduce a dramatic increase in multipath scattering, which could otherwise degrade the channel performance. This is a positive finding, as it suggests that THz systems can perform adequately even during rain, with the direct path only suffers power loss mainly due to absorption by rain [28] and can be compensated by employing some power control strategies, or other environmental adaptation techniques. We think the reason for this phenomenon is that, although raindrops cause scattering, the absorption loss could be much more higher [30] and most of the scattered multipath components is absorbed or dissipated before reaching the receiver. This leads us to conclude that inter-symbol interference (ISI) caused by multipath scattering rarely occurs under rainy conditions.

In the above analysis, we employed the averaged K-factor and SNR, for several groups of measurements conducted within a very short time interval (typically within 1-2 minutes) under identical or very similar rain rates (variation less than 0.5 mm/hr), to isolate the randomness corresponding to the variation of rain rate and rain drop size distribution (RDSD). Here, we further concentrate on the variation of K-factor and power variance for each time of measurement. As shown in Fig. 3(a), the K-factor generally decreases with increasing rain rate, indicating a weakening of the LOS component and the emergence of more scattered signals. However, an important observation is that, for the same or nearly similar rain rates, the K-factor still shows some variability. This suggests that even under close or identical rain rates, other environmental factors (such as random variations in raindrop size and distribution) could influence the balance between the LOS and scattered signal components. This phenomenon is consistent with the observation of power variance in Fig. 3(b), where similar rain rates sometimes result in different levels of power variance. Similar phenomenon has been observed for millimeter wave channels [32], and might make channel quality degrade suddenly and bit error rate spikes unexpectedly.

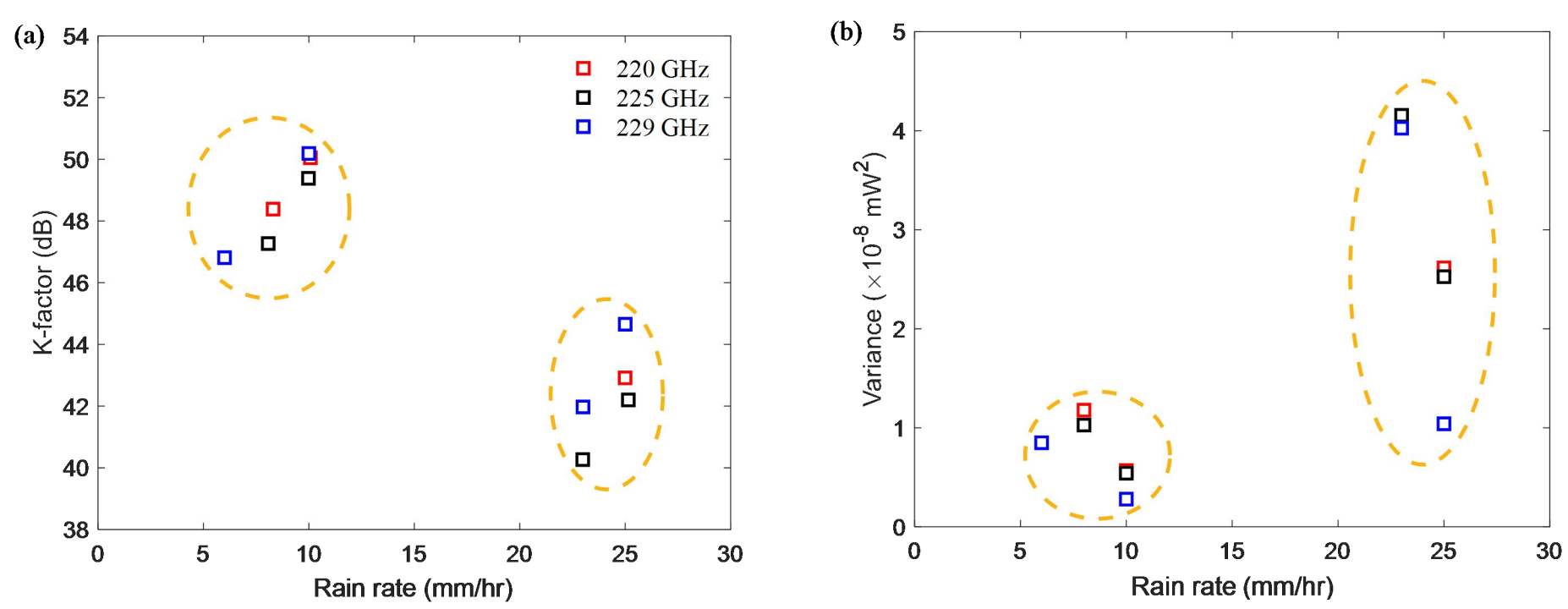

**Fig. 3.** Performance of (a) K-factor and (b) variance under different rain rates. Both (a) and (b) keep the same legend.

### *3.2. Theoretical modeling*

To achieve precise modeling of terahertz channels in rainy environments, we referred to the ITU-R P.838-3 recommendation [43] and optimized the original ITU model by applying an empirical fitting method. The formula used in this work is

$$\gamma_R = kR^{\alpha} \tag{1}$$

with R representing the rain rate in mm/hr. $k$ and $\alpha$ are empirical parameters derived from the ITU recommendations and tailored for different channel measurement scenarios based on horizontal and vertical polarization. For our measurement setup, we employed the empirical parameters for horizontal polarization. These parameters were obtained by fitting curves to power-law coefficients derived from discrete calculations based on empirical data. At 220 GHz, the values are k=1.63596 and α=0.63648 at 225 GHz, k=1.63504 and α=0.63562; and at 229 GHz, k=1.63412 and α=0.63476.

Another method that employed is the Mie scattering theory, as

$$\alpha = 3.3429\int_0^\infty p(r)\xi_e(r)\cdot \pi r^2 dr \qquad (2)$$

with the p(r) being the RDSD depending on the radius of rain droplets and $\xi_e(\chi)$ being the extinction efficiency. There are a number of RDSD models for choice, including M-P [44], Joss-Waldvogel (J-W) [45], Joss-Drizzle(J-D) [46], Joss-Thunderstorm (J-T) [46]. The M-P distribution model is typically used to describe raindrop size distributions in moderate (2.5 mm/hr ≤ R < 10 mm/hr) to heavy rainfall (10 mm/hr ≤ R < 50 mm/hr) [44], while the J-W, J-D, and J-T models are generally applied to describe raindrop size distributions in heavy rain, drizzle, and thunderstorm-type rainfall, respectively [47, 48]. All four distribution models can be expressed in the form of a negative exponential function $N(D) = N_0 exp(-\Lambda D)$, where D refers to the raindrop diameter in mm. $N(D)$ is the number density of raindrops of diameter D in a unit volume. The corresponding parameters shown in Table 3. In this model, we used the dielectric properties of pure water calculate by using the double Debye dielectric model (D3M) [49], which incorporates empirically fitted parameters and is particularly suitable for high-frequency electromagnetic waves, such as those in the 220 GHz-230 GHz range used in this work. Here we do not include the influence of gaseous absorption, because the variation of temperature and humidity in clear and rainy weather conditions was not obvious.

**Table 3. Raindrop Size Distributions and Their Corresponding Parameter [44-46, 50]**

| Distribution model | $N_0$ $(m^{-3}mm^{-1})$ | $\Lambda$ $(mm^{-1})$ |
|---|---|---|
| M-P | 8000 | $4.1R^{-0.21}$ |
| J-W | 7000 | $4.1R^{-0.21}$ |
| J-D | 30000 | $5.7R^{-0.21}$ |
| J-T | 1400 | $3.0R^{-0.21}$ |

The calculation results are shown in Fig. 4. Through systematic comparison of different theoretical approaches, we found that in moderate rain (R < 10 mm/hr), calculation by Mie scattering theory fits the measured data better. However, when the rainfall rate increases to heavy rain (R ≥ 10 mm/hr) conditions, most of the measured data fall between the predictions by the ITU model and the J-D model. Notably, among all Mie scattering models examined, the J-D model demonstrates superior performance by providing the largest rain attenuation prediction, becoming a reliable lower bound for predicting channel power loss in rain. The ITU-R P.838-3 model tends to overestimate attenuation caused by rainfall, providing a higher bound for estimation of power loss by rain.

One notable phenomenon observed in the data is the variability in power loss at identical rain rates. This variability (consistent with the observation in Fig. 3) implies that, even under constant rain intensity, other influencing factors are causing fluctuations in signal attenuation.

Unlike the controlled rainfall environment in some indoor emulated experiments [31, 51], in natural settings, rainfall intensity is inherently variable, making it difficult to maintain stable conditions even during short-term rain rate measurements. Outdoor rainfall environments are also characterized by randomness, with factors such as rain intensity and RDSD, both of which can constantly change and affect THz channel performance. This has been confirmed in previous studies [27, 52], where significant differences in RDSD have been observed in different spatial regions of the transmission channel. Even at the same rainfall rate, variations in raindrop size and total quantity lead to noticeable differences in rain attenuation [53]. Additionally, variations in environmental factors such as wind speed, humidity, and temperature in the rainfall experiment environment can sometimes affect the channel [15, 54], even though they are measured almost identical in this work. This is consistent with our observations in Fig. 3(b) and presents a significant challenge for accurate channel modeling, which assume uniform RDSD, and can result in communication failures or degraded performance. Given these inherent variations and the complementary strengths of different models, we selected both the ITU model and the J-D model to establish a prediction range for channel power attenuation rather than attempting to predict exact values, as this approach better captures the natural variability in terahertz channel performance under rainfall conditions.

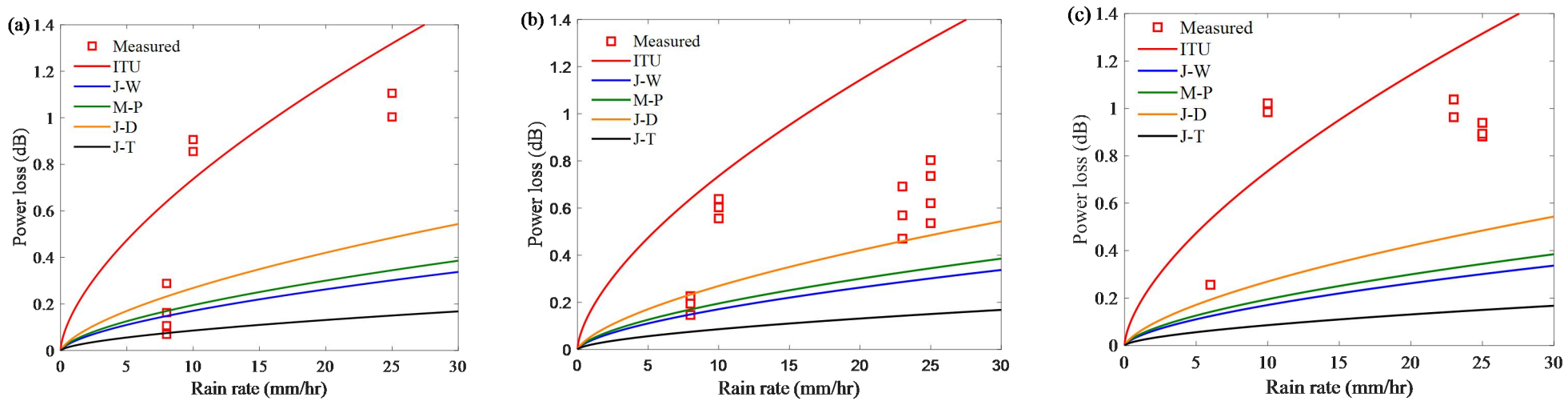

**Fig. 4.** Channel power loss as a function of rain rate for the (a) 220 GHz, (b) 225 GHz, and (c) 229 GHz channels over a 41.5 m channel path in rain under experimental measurement and theoretical prediction.

## 4. Channel BER performance

The preceding theoretical model illuminates channel power profiles. However, translating these findings into a comprehensive understanding of communication link reliability requires a deeper exploration. To this end, BER analysis becomes an indispensable tool, offering critical insights into how power variations influence the error rates in transmitted data.

Our analysis includes several digital modulation schemes, such as Binary Phase Shift Keying (BPSK), Frequency Shift Keying (FSK), Pulse Amplitude Modulation (PAM), and Quadrature Amplitude Modulation (QAM), which cover a wide range of modulation complexities and sensitivity to different types of channel impairments. As identified in previous studies (see Table 2), the Rician model fits the measured values better and the Rician fading channel models are employed in this work. The K-factor used in the analysis is the average K-factor value of 46 dB obtained from the measurements. The range of the energy per bit to noise power spectral density ratio ($E_b/N_0$) is set between 0 dB and 20 dB. The calculation results by using the 'bertool' function in MATLAB R2024a are shown in Fig. 5, by using BPSK, 2-FSK, and 2-PAM, as well as 4-QAM modulation schemes. The BER performance under 2-FSK is worst, due to its sensitivity to frequency-dependent attenuation and its inefficient use of energy. In contrast, BPSK, 2-PAM, and 4-QAM are less sensitive to these issues because they modulate amplitude and/or phase rather than frequency, making them more robust against such channel impairments. It should also be noted that BPSK, 2-PAM, and 4-QAM offer the same BER performance, despite 4-QAM being a higher-order modulation scheme (4-ary) compared to the lower-order (2-ary) BPSK and 2-PAM. We think this is because the channel conditions (e.g., Rician fading with a strong LOS component) provides a relatively stable propagation environment. In such a channel, the effects of multipath fading are minimized, and the channel experiences less distortion. This stability allows higher-order modulation schemes to maintain performance comparable to simpler schemes. This makes us to take the QAM modulation for our following calculations, together with higher-order modulation schemes to achieve greater transmission efficiency. Besides, QAM modulation supports various advanced signal processing technologies, such as MIMO, OFDM, and adaptive modulation.

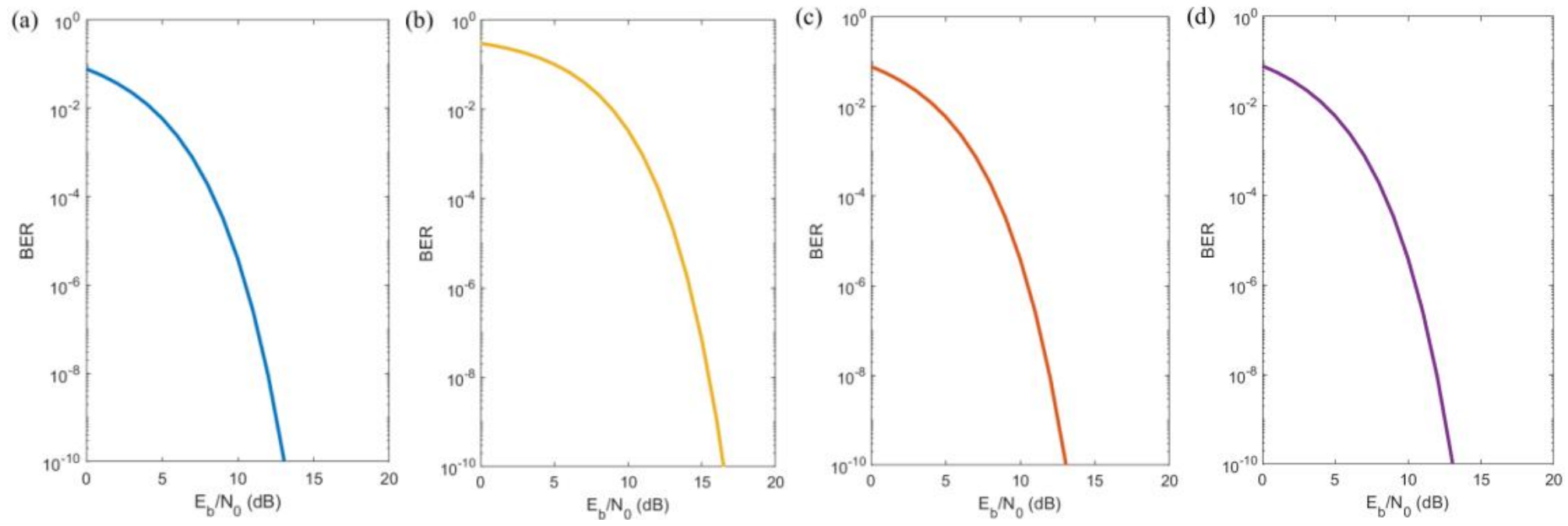

**Fig. 5.** BER performance under different modulation schemes as (a) BPSK modulation, (b) 2-FSK modulation, (c) 2-PAM modulation, (d) 4-QAM.

We take the 16-QAM modulation into consideration, as it has been employed in several publications for high-quality data transmission with high spectral efficiency [13, 55], and can be updated to much higher orders [56]. By employing the 16-QAM modulation scheme, we could comprehensively analyze the effects of different rain rates, temperature, and humidity on THz channels in rainy environments. We used the BER formula for QAM modulation under Rician fading channels as

$$BER_{\text{16-QAM}} = \int_0^{\infty} \frac{3}{8}\left(1 - \sqrt{\frac{\gamma}{5+\gamma}}\right) f(\gamma)\, d\gamma \tag{3}$$

where $\gamma$ represents the signal-to-noise ratio (SNR). $f(\gamma)$ is the probability density function (PDF) of SNR, given by the following the formula

$$f(\gamma) = \frac{1+K}{\bar{\gamma}} \exp\left(-K - \frac{(1+K)\gamma}{\bar{\gamma}}\right) I_0\left(2\sqrt{\frac{K(1+K)\gamma}{\bar{\gamma}}}\right) \tag{4}$$

with $\bar{\gamma}$ being the average SNR and $I_0(\cdot)$ as the modified Bessel function of the first kind and order zero. We proposed a formula as $K(d) = K_0 \cdot (d_0/d)^{\alpha}$ to calculate the K-factor, which varies with respect to channel distance (d), and path loss exponent (α) depending to channel power as $\alpha = \left[\bar{P}_L(d) - \bar{P}_L(d_0)\right] / \left[10\log(d/d_0)\right]$ [57]. Here, $d_0$ was taken as 41.5 m for reference and $\bar{P}_L(d_0)$ and $\bar{P}_L(d)$ represent the power loss values at distances $d_0$ and $d$, respectively. Based on this, by incorporating the rain rate into the K-factor calculation, we obtain the formula for the K-factor as

$$K = K_0 \cdot \left(\frac{d_0}{d}\right)^{\alpha} \cdot \left(1 - \frac{R}{R_c}\right) \tag{5}$$

where, $R$ represents the rain rate in mm/hr, and $R_c$ is the critical rain rate, defined as the threshold above which significant attenuation occurs. It was set to be $R_c$ = 50 mm/hr in this work, as the rain intensity reaches to be extreme [58]. We calculate the BER performance by employing the ITU model for the upper bound and the J-D model for the lower bound. We think the BER values should be in the covering area. As shown in Fig. 6, if the channel is launched out with a transmitting power of 20 dBm over a channel distance of 41.5 m, low bit error transmission (BER ~ $2\times10^{-7}$) keeps unchanged as the rain rate increases up to 50 mm/hr. This means the channel would not be affected when it propagates over a short distance, where the total accumulated attenuation caused by rain over such a short distance isn't enough to significantly degrade the channel quality. However, when we extend the transmission distance to be 1 km, significant BER degradation can be observed and the channel fails as the BER is always above the forward error correction (FEC) threshold [14], making reliable communication impossible. When we increase the transmitted power to be 40 dBm, which is a large radiation power for THz waves but still achievable [59], the BER below the FEC threshold can be achieved under rain rate R < 30 mm/hr. This indicates the increasing transmit power is a viable strategy for maintaining channel performance in rain, even though this could increase the hardware complexity and reduce power efficiency.

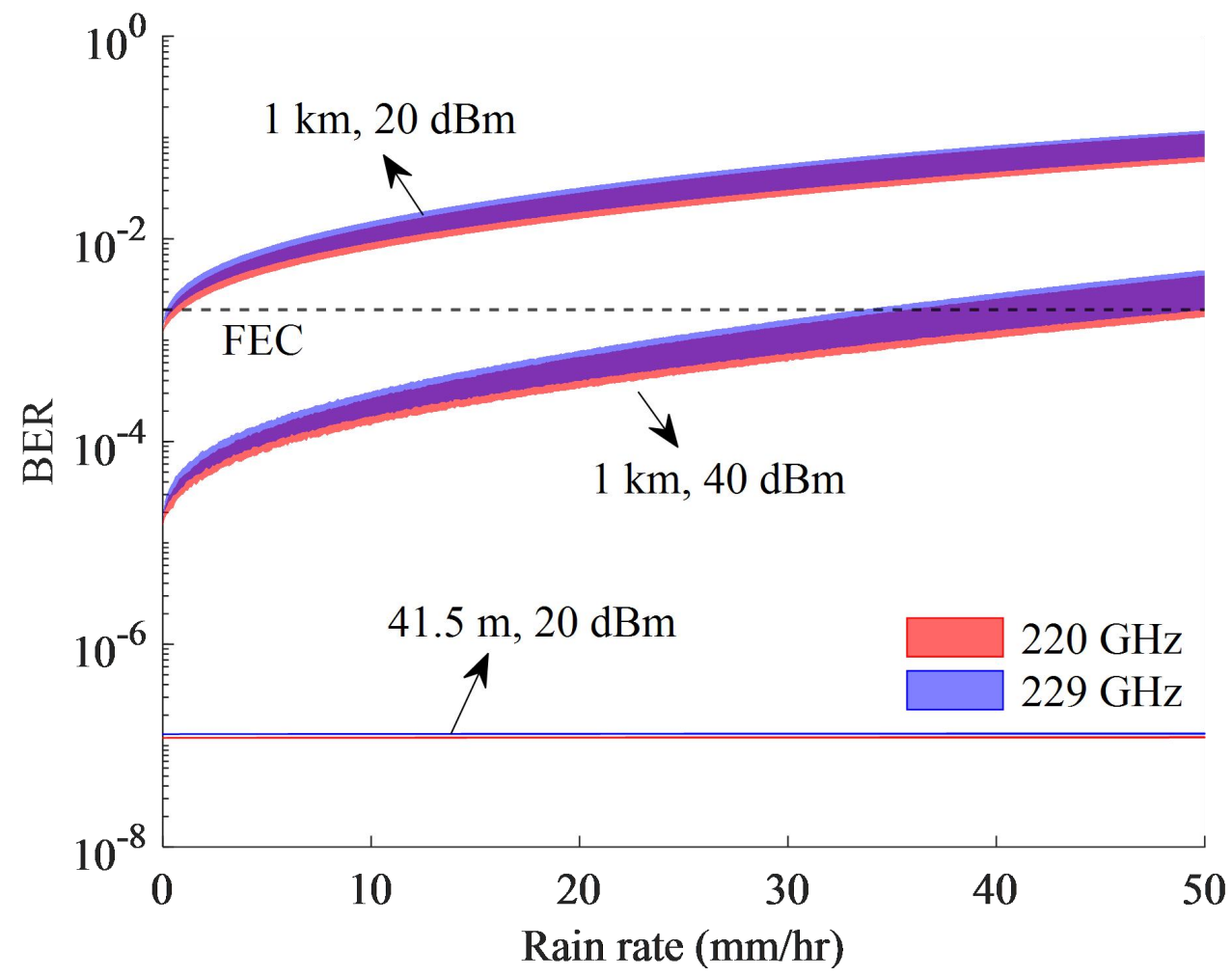

**Fig. 6.** Predicted BER performance for channels operating at 220 GHz, 225 GHz and 229 GHz. The upper and lower bounds of the prediction region correspond to the predicted values of the ITU model and the J-D model, respectively. Channel distance 1 km, relative humidity RH 82%, temperature 27°C, transmit power 20 dBm, noise level of receiver -60 dBm; the gain at the transmitter side is 60.5 dB (combination of antenna, lens and power amplifier) and at the receiver side is 40.5 dB (combination of antenna and lens).

To further investigate the BER performance of terahertz channels under different rainy conditions, we predict the BER variation due to changes of rain rate, temperature, and relative humidity. The ambient temperature is always set to 27°C, and the relative humidity is always 82%, as observed during our experimental measurements. An AWGN channel was employed for the channel propagation in clear weather for comparison. The BER performance was calculated by using Eq. (3) for the channel in rain and using the following formula for the channel in clear weather, as

$$BER_{16-QAM,AWGN} \approx \frac{3}{8} \cdot Q(\sqrt{\frac{4}{5}\gamma}) \tag{6}$$

with $Q(x)$ being the Q-function. It can be seen from Fig. 7(a) that, under 16-QAM modulation, the operational frequency of terahertz channel in clear weather is limited to 350 GHz, due to the limitations of gaseous absorption peaks. In rainy conditions, the available threshold drops to 300 GHz, as the increasing of the total power loss caused by rain. It is evident that as the rain rate increases from 5 mm/hr to 25 mm/hr, there is a significant decline in channel performance. However, within this range of rain rates, there still exists operational frequency (< 250 GHz) and corresponding frequency band allowing the channel to function normally, and frequency planning should focus on lower THz bands to reduce sensitivity to rain at higher rain rates. It should also be note that the presence of Rician fading introduces fluctuations of the predicted BER curves, but this impact it not so obvious and can be negligible. This confirms our previous conclusion that the ISI caused by multipath scattering is negligible under rainy conditions.

To further investigate the influence of temperature and humidity, which can vary significantly across different geographical regions and seasons during rainfall conditions, we plot the BER performance with respect to both parameters, as shown in Fig. 7(b) and (c). Here, we choose the carrier frequency at 140 GHz and 220 GHz to avoid the gaseous absorption peaks as indicated in Fig. 7(a), where the channel works with low BER in rainfall. As shown in Fig. 7(b), with the increasing of temperature, the BER increases, as temperature affects the dielectric constant of rain droplets according to the double Debye dielectric model (D3M) [49]. However, the BER performance shows a trend of slight variation with temperature. This means temperature appears less critical compared to rain rate, as it varies over a short and even negligible range. In Fig. 7(c), BER increases more noticeably with rising relative humidity, which makes humidity as a pronounced impact on the reliability of THz communication systems, not only in rainy weathers. It should also be noted that the 220 GHz channel shows greater sensitivity to humidity changes, and the 140 GHz channel shows better resilience to humidity variations. Thus, the selection of operating frequency for THz communication systems should carefully weigh the tradeoffs between higher frequencies' increased data capacity versus their greater environmental sensitivity when selecting operating frequencies for different deployment scenarios.

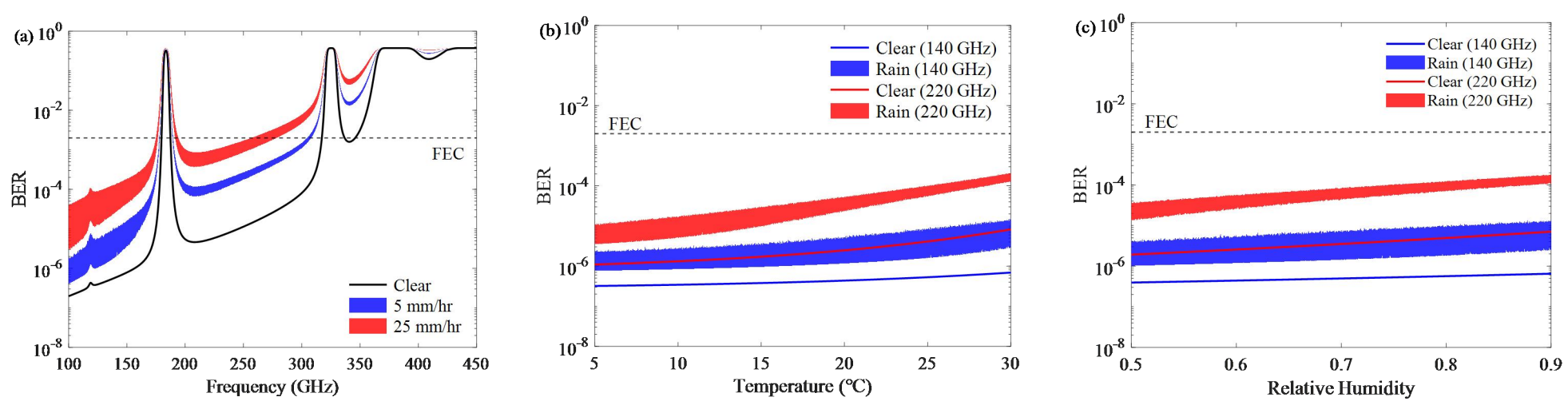

**Fig. 7.**(a) Variation of channel BER with frequency at different rainfall rates 5 mm/hr and 25 mm/hr and in clear weather. (b) Variation of channel BER with temperature. (c) Variation of channel BER with relative humidity. Channel distance 1 km, humidity RH 82%, temperature 27°C, linear K-factor is calculated by for Eq. (5), transmitted power 40 dBm, noise level of receiver -60 dBm, the gain at the transmitter is 60.5 dB (combination of antenna, lens and power amplifier) and receiver side is 40.5 dB (combination of antenna and lens). In both Fig. 7.(b) and Fig. 7.(c), the rain rate is set to 5 mm/hr.

## 5. Conclusion

In this article, we conducted a systematic investigation of terahertz channel behavior under rainfall conditions, employing both outdoor measurements and sophisticated theoretical modeling approaches. Our experimental campaign, conducted at the Beijing Institute of Technology, utilized a 54-meter outdoor channel operating at frequencies between 220-230 GHz, enabling detailed characterization of power profiles and BER performance under various rainfall conditions. Analysis of the CDF profiles revealed consistent conformity to Rician distributions with K-factors maintaining values above 40 dB even in heavy rainfall, indicating minimal multipath scattering effects. Notably, we observed significant variations in power loss under constant rain rates, attributable to the dynamic nature of raindrop size distributions and their complex interactions with THz waves.

In terms of theoretical modeling, we found that while the ITU-R model tended to overestimate attenuation, the Mie scattering approach using the J-D distribution provided a reliable lower bound for predicting channel power loss in rain. Our comprehensive BER analysis demonstrated the superior performance of QAM compared to other modulation schemes, with an important finding that channels operating at higher frequencies exhibited increased sensitivity to humidity variations. This frequency-dependent behavior suggests that lower frequency channels offer greater environmental resilience, though at the cost of reduced data capacity.

Several limitations in our current study present opportunities for future research. While our measurements captured overall power attenuation effectively, the precise characterization of rapid temporal variations in channel properties under changing rainfall conditions remains challenging. Additionally, the impact of varying raindrop size distributions on channel coherence time and phase stability requires further investigation, particularly at higher frequencies where these effects may become more pronounced. Future work should focus on developing more sophisticated channel models incorporating dynamic raindrop size distributions, investigating adaptive modulation techniques for maintaining reliable communications during adverse weather conditions, and exploring advanced signal processing algorithms specifically designed for rain-induced channel impairments.

**Funding.** National Natural Science Foundation of China under Grant (62471033, 92167204, 62072030), Science and Technology Innovation Program of Beijing Institute of Technology (2022CX01023), Fundamental Research Funds for the Central Universities (2024CX06099) and the Talent Support Program of Beijing Institute of Technology "Special Young Scholars" (3050011182153).

**Disclosures.** The authors declare no conflicts of interest.

**Data availability.** Data underlying the results presented in this paper are not publicly available at this time but maybe obtained from the corresponding author - Jianjun Ma upon reasonable request.